\documentclass[aps,twocolumn]{revtex4}
\usepackage{amsfonts}

\def\half{\textstyle{\frac{1}{2}}}

\def\bb{\beta}
\def\t{\theta}

\def\ra{\rightarrow}
\def\tint{{\textstyle\int}}

\def\s{\hskip.08em}
\def\ss{\sigma}
\def\d{\partial}

\def\b{\begin{eqnarray}}  
\def\e{\end{eqnarray}}    
\def\bn{\begin{eqnarray}}  

\def\<{\langle}
\def\>{\rangle}

\def\no{\nonumber}
\def\Tr{{\rm Tr\s}}

\def\{{\lbrace}
\def\}{\rbrace}
\begin{document}
%
%
%
\title{Generalized Phase-Space Representation of Operators}
\author{John R.~Klauder$^{a,b,}$}
  \email{klauder@phys.ufl.edu}
\author  {Bo-Sture K. Skagerstam$^{c,}$}
  \email{boskag@phys.ntnu.no}

\affiliation
  {\vspace{1mm} \small  $^a$Department of Physics and Department of Mathematics,
                           University of Florida,
                           Gainesville, FL 32611, U.S.A.\footnote{Permanent address}\\
$^b$Department of Physics,
The Norwegian University of Science and Technology,
 N-7491 Trondheim,
Norway\\
$^c$Complex Systems and Soft Materials Research Group, Department of Physics,
The Norwegian University of Science and Technology,
N-7491 Trondheim, Norway}
\begin{abstract}
\noindent Introducing asymmetry into the Weyl representation of
operators leads to a variety of phase-space representations and new
symbols. Specific generalizations of the Husimi and the
Glauber-Sudarshan
symbols are explicitly derived.\\
\\
\noindent PACS numbers 42.50.-p, 42.25.Kb, 03.65.-w
     $~~~~~~~~~~~~~~~~~$ \today
\end{abstract}
\maketitle
\section{Introduction}
Phase space representations have proven their value in classical
mechanics and ever since the work of Wigner \cite{wig} in quantum
mechanics as well. The map introduced by Wigner from state functions
to quasi-probability functions is notable for two fundamental
properties. The first reason it is notable relates to the fact that,
in general, the Wigner quasi-probability function is not everywhere
nonnegative, and this keeps it from being a true probability
function. The second reason it is notable relates to the fact that
the Wigner map is not unique, and this property has led to a variety
of alternative prescriptions to define a quantum mechanical phase
space distribution. The most widely known variation on the Wigner
map is that due to Cohen \cite{cohen}. Cohen's generalization, which
is typical of most such efforts, normally steps outside the family
of operator images by the Wigner map; in other words,
generalizations are introduced that involve an auxiliary phase-space
function that is generally not obtained by means of an operator map.

Two well known phase-space representations for operators, however,
are defined by direct transcriptions of operators. In particular, we
refer to the Husimi phase-space representation, and its dual, the
Glauber-Sudarshan phase-space representation, both of which are
recalled below. Both representations can be defined in an entirely
abstract operator setting with the only transcription between
operators and phase space being given by an exclusive use of the
Weyl operator. The goal of the present paper is the generalization
of such schemes leading to a multitude of such dual pairs that are
defined in a strictly operator fashion, and which, thereby,
transform covariantly under coordinate transformations.

While our discussion is presented in the notation and language of
quantum mechanics, a completely parallel discussion applies to a
time-frequency analysis of signals and their transformations; in
this connection, see \cite{charlie}, Sec.~4.5 and Chap.~9.

\subsection{Basics: Wigner, Weyl, and Moyal} The basic phase space
representation of quantum states and operators is due to Wigner
\cite{wig}, Weyl \cite{weyl} and Moyal \cite{moyal}. The basis of
this representation involves the Heisenberg canonical operator pair,
$P$ and $Q$, subject to the commutation rule $[Q,P]=i$ (with
$\hbar=1$ throughout), expressed in their exponential form
    \b U[p,q]\equiv e^{i(pQ-qP)} \;, \e
    and known as the Weyl operator.
    As such, we assume the operators $Q$ and $P$ are both self
    adjoint, with continuous spectrum on the entire real line,
     and thus the operators $U[p,q]$ are unitary and
    (weakly)
    continuous in the real parameters $(p,q)\in {\mathbb R}^2$.
    In terms of these operators the commutation relation assumes the form
    \b
    U[p,q]\,U[r,s]&=&e^{i(ps-qr)/2}\,U[p+r,q+s]\nonumber \\&=&e^{i(ps-qr)}\,U[r,s]\,U[p,q]\;,
    \label{u9}\e
    which is the version we shall use. A key relation we shall
    employ is the distributional identity
    \b \Tr(U[p,q]\, U[r,s]^\dag)&=&\Tr( U[r,s]^\dag\,U[p,q])\nonumber\\ &=&2\pi\,\delta(p-r)\s\delta(q-s)\;, \e
    and it is appropriate that we reestablish this well known expression.
    To that end, consider the diagonal position space matrix elements
    \b && \hskip1.00cm \<y|\s U[p,q]\,U[r,s]^\dag\s|y\> \nonumber\\ &=&\<y|\s e^{ipQ/2}\s e^{-iqP}\s
    e^{i(p-r)Q/2}\s e^{isP}\s e^{-irQ/2}\s|y\> \nonumber\\
&=&e^{i(p-r)y/2}\<y|\s e^{-iqP}\s e^{i(p-r)Q/2}\s
    e^{isP}\s|y\>\nonumber\\
&=&e^{i(p-r)y/2}\<y+q|\s e^{i(p-r)Q/2}\s|y+s\>\nonumber \\
&=&e^{i(p-r)y/2}\,e^{i(p-r)(y+s)/2}\,\delta(q-s)\;. \e
Integration over $y$ leads to
 \b \label{eq:basic} \Tr(U[p,q]\, U[r,s]^\dag) &=&\int
 e^{i(p-r)y}\,e^{i(p-r)s/2}\,\delta(q-s)\,dy
\nonumber\\ &=&2\pi\,\delta(p-r)\,\delta(q-s)\;, \e as required. We
make occasional use of this identity, which we call the {\it basic
identity}.

The Weyl representation of operators (limited to Hilbert-Schmidt
operators for the time being), is given by
   \b A=\int U[k,x] \,{\tilde A}[k,x]\,\frac{dk\s
   dx}{2\pi} \;, \e
   where ${\tilde A}[k,x]$ is a function to be determined.
   If we consider $\Tr(U[k',x']^\dag\, A)$ and use the
   basic identity Eq.~(\ref{eq:basic}), we learn that
   \b  \Tr(U[k',x']^\dag\,A)&=&\int \Tr(U[k',x']^\dag\,U[k,x])\,{\tilde
   A}[k,x]\frac{dk\s
   dx}{2\pi} \nonumber\\
&=&{\tilde A}[k',x']\;.  \e
   Therefore, we have the important completeness relation
   \b \label{eq:complete} A=\int U[k,x]
   \,\Tr(U[k,x]^\dag\,A)\,\frac{dk\s
   dx}{2\pi} \;. \e
We take a second operator with an analogous representation,
     \b B=\int U[k,x] \,\Tr(U[k,x]^\dag\,B)\,\frac{dk\s
   dx}{2\pi} \;,
      \e
      and we again introduce the notation that
      \b {\tilde B}[k,x]\equiv\Tr(U[k,x]^\dag\,B)\;. \e
     Using the basic identity Eq.(\ref{eq:basic}) again, we learn that
     \b &&\hskip2.00cm \Tr(A^\dag\,B)\nonumber \\ &=& \int{\tilde
     A}[k',x']^*\Tr(U[k',x']^\dag\,U[k,x]){\tilde B}[k,x]\,\frac{dk'
     dx'\s dk\s dx}{(2\pi)^2}\nonumber \\
&=&\int{\tilde A}[k,x]^*\,{\tilde B}[k,x]\,\frac{dk\s dx}{2\pi}
\;.\label{w15} \e

In summary, the Weyl representation for operators is given by
   \b A=\int U[k,x] \,{\tilde A}[k,x]\,\frac{dk\s dx}{2\pi} \;,\e
   where
    \b {\tilde A}[k,x]\equiv\Tr(U[k,x]^\dag\,A)\;. \e
These relations are complemented by the expression
  \b\Tr(A^\dag\,B)=\int{\tilde A}[k,x]^*\,{\tilde B}[k,x]\,\frac{dk\s dx}{2\pi}\;.\e

  It will also be important to introduce the Fourier transforms of
  the Weyl representation elements as
   \b A[p,q]\equiv \int e^{i(kq-xp)}\,{\tilde A}[k,x]\,\frac{dk\s dx}{2\pi}
   \;, \e
   and, correspondingly,
   \b B[p,q]\equiv \int e^{i(kq-xp)}\,{\tilde B}[k,x]\,\frac{dk\s dx}{2\pi}
   \;.\e
   In terms of these representatives we have
   \b\Tr(A^\dag\,B)=\int{A}[p,q]^*\,{B}[p,q]\,\frac{dp\s
  dq}{2\pi}\;,\label{w17}\e
  as follows from Parseval's Theorem.

   We note that $A[p,q]$ and
  $B[p,q]$ as introduced here, are generally referred to as
  the Weyl symbols for the operators $A$ and $B$, respectively.
  Hence, ${\tilde A}[k,x]$ and ${\tilde B}[k,x]$ are the
  Fourier transform of the corresponding Weyl symbols.

\section{Initial Introduction of Asymmetry}

It is clear from Eqs.~(\ref{w15}) and (\ref{w17}) that the Weyl
representation has led to a {\it symmetric} functional realization
in how it treats the operators $A$ and $B$. This is a natural
representation choice for applications in which $A$ and $B$ enter in
a symmetric manner. However, there are cases when that is not
appropriate, and in such situations it can prove valuable to treat
the {\it representations} for $A$ and $B$ in an {\it asymmetric}
manner. For example, it is well known that in quantum optics
observables are commonly represented by normally ordered operators,
which are readily given a phase space representation via the Husimi
 representation \cite {husimi}. In turn, the dual representation for
this important example is known as the Glauber-Sudarshan
representation
 \cite{glasud}. We begin our analysis of asymmetric representations
 with this important conjugate pair.

To discuss these alternative representations it is convenient to
first recall canonical coherent states given by
  \b |p,q\>\equiv U[p,q]\s|\s 0\>\;, \e
  where the fiducial vector $|\s 0\>$ is the normalized ground state of a suitable
  harmonic oscillator, or more simply that
    \b (Q+iP)\,|\s 0\>=0 \;. \e

    From a notational point of view, it is also useful to recall
    that
    \b |p,q\>&\equiv&|z\>=e^{za^\dag-z^*a}\s|\s 0\>
=e^{-|z|^2/2}\,e^{za^\dag}\,e^{-z^*a}\s|\s 0\>\nonumber \\
&=&e^{-|z|^2/2}\,e^{za^\dag}\s|\s 0\>
=e^{-|z|^2/2}\sum_{n=0}^\infty \frac{z^n a^{\dag\s n}}{n!}\s|\s 0\>\nonumber\\
&=&e^{-|z|^2/2}\sum_{n=0}^\infty\frac{z^n}{\sqrt{n!}}\s|\s
    n\>\;.
    \e
    In this expression we have introduced $z=(q+ip)/\sqrt{2}$,
    $a=(Q+iP)/\sqrt{2}$, from which it follows that $[a,a^\dag]=1$ and $a\s|\s 0\>=0$,
    and finally the normalized states $|\s n\>\equiv a^{\dag\s n}\s|\s
    0\>/\sqrt{n!}$ for all $n\ge0$. These are the familiar canonical
    coherent states; however, since we will generalize expressions
    well beyond these particular states, we shall not focus on the
    complex parametrization which is generally limited to this
    special family of coherent states.

Returning to the phase space notation, the Husimi phase-space
representation of $B$ is given by
     \b B_H[p,q]&\equiv& \<p,q|\s B(P,Q)\s|p,q\>\nonumber\\&=&\<0|\s B(P+p,Q+q)\s|\s 0\>\;;  \e
     and the operator representation given by
     \b A\equiv \int A_{G-S}[p,q]\,|p,q\>\s\<p,q|\,\frac{dp\s
  dq}{2\pi}\;, \label{q3}
     \e
     involves the Glauber-Sudarshan weight function $A_{G-S}[p,q]$
     in a weighted integral over the one dimensional
     projection operators $|p,q\>\s\<p,q|$.
     It is clear that the two symbols $A_{G-S}[p,q]$ and $B_H[p,q]$
     are related to their respective operators (i.e., $A$ and $B$)
     in very different ways. Nevertheless, one has the important relation
     \b \Tr(A^\dag \,B)&=&\int A_{G-S}[p,q]^*
     \,\Tr(|p,q\>\s\<p,q|\,B)\,\frac{dp\s
  dq}{2\pi}\nonumber\\
&=&\int A_{G-S}[p,q]^*\,B_H[p,q]\,\frac{dp\s
  dq}{2\pi}\;, \e
     and it is in this sense that we refer to these two distinct
     representations as dual to one another.

     Despite their very different appearances, there is a fairly
     close
     connection between the initial, symmetric Weyl representations
     and the latter, asymmetric Husimi -- Glauber-Sudarshan representations.
     That relationship is most easily seen in the Fourier
     transforms of the given symbols.
     First, let us start from the symmetric Weyl expression
     (\ref{w15})
     and modify that expression in the following way:
       \b &&\hskip2.00cm \Tr(A^\dag\,B)\nonumber\\ &=&\int \{\s e^{(k^2+x^2)/4}
       \s{\tilde A}[k,x]\}^*\,\{\s e^{-(k^2+x^2)/4}\s{\tilde
       B}[k,x]\}\,\frac{dk\s dx}{2\pi}\,\nonumber \\
&\equiv&\int {\tilde A}_{G-S}[k,x]^*\,{\tilde B}_H[k,x]\,\frac{dk\s
dx}{2\pi}\;.\label{u5}\e
      In (\ref{u5}) we have introduced
          \b {\tilde A}_{G-S}[k,x]=e^{(k^2+x^2)/4}\s{\tilde
          A}[k,x]\nonumber \\
=\int e^{i(px-qk)}\,A_{G-S}[p,q]\,dp\s dq/2\pi\;, \e
          and
     \b {\tilde B}_H[k,x]&=&e^{-(k^2+x^2)/4}\s{\tilde B}[k,x]\nonumber \\
&=&\int e^{i(px-qk)}\,B_H[p,q]\,dp\s dq/2\pi\;.  \e
     This connection between the Weyl and the Husimi -- Glauber-Sudarshan representations
     has been known for some time (see, e.g.,
     Ref.~\cite{klasud}, p. 185), and we do not repeat a proof here; it
     will be implicitly reestablished in what follows.

     A direct connection also exists between the Husimi and
     the Glauber-Sudarshan representations as well. In particular,
     we can transform Eq.~(\ref{q3}) to read
     \b  \hskip-.4cm A_H[r,s]&=&\int A_{G-S}[p,q]\s\,|\<p,q|r,s\>|^2\,\frac{dp\s
     dq}{2\pi} \no\\
       &=&\int A_{G-S}[p,q]\,e^{-[(r-p)^2+(s-q)^2]/2}\,\frac{dp\s
     dq}{2\pi}\,. \label{q8} \e
     The functional form of this equation as a convolution reflects
     the connection through multiplication in the Fourier space.

     It is clear from the connections above that the Husimi symbol $B_H[p,q]$ is generally
     ``smoother'' than the corresponding Weyl symbol $B[p,q]$, while the Glauber-Sudarshan symbol
     $A_{G-S}[p,q]$ is generally ``rougher'' than the corresponding
     Weyl symbol $A[p,q]$. In any asymmetric treatment of the
     representations this dichotomy is inevitable. In particular, it
     is known (see Ref.~\cite{klasud}, p. 183) that for bounded operators,
     for example, the Glauber-Sudarshan
     symbol $A_{G-S}[p,q]\in {\cal Z}'({\mathbb R}^2)$, which is the
     Fourier transform of the more familiar space  of
     distributions in two dimensions, ${\cal D}'({\mathbb R}^2)$. The use of suitable distributions
     for phase-space symbols is perfectly acceptable provided that
     they are always paired with dual symbols that lie in
     the appropriate test function space.
     \section{Generalized Asymmetric Phase Space Representations}
     We now come to the main topic of this paper, namely the
     introduction of a large class of asymmetric phase space
     representations. In so doing, we shall see how this analysis
     incorporates and generalizes the familiar asymmetric example
     discussed in the previous section.

     Let us introduce a nonvanishing operator $\sigma$ which we require to be
     a trace-class operator, i.e., we require that $0<\Tr(
     \sqrt{\sigma^\dag\s\sigma}\s)<\infty$. Such operators have the
     generic form given by
        \b \sigma=\sum_{l=0}^\infty c_l\,|\s b_l\>\s\<a_l|\;, \e
        where $\{|\s a_l\>\}_{l=0}^\infty$ and $\{|\s b_l\>\}_{l=0}^\infty$
        denote two, possibly identical, complete orthonormal sets of vectors, and the
        coefficients $\{c_l\}_{l=0}^\infty$ satisfy the condition
          \b \Tr(\sqrt{\ss^\dag\s\ss}\s)=\sum_{l=0}^\infty |c_l|<\infty\;. \e
       If $\sigma^\dag=\sigma$, we may choose  $|\s b_l\>=|\s a_l\>$,
       and the coefficients $c_l$ as real for
       all $l$; however, it is not
       required that $\ss$ be Hermitian.

       We shall have need of the function $\Tr(U[k,x]\s\ss)$ defined for all
       $(k,x)$ in phase space. Observe that because $\ss$ is trace class, this function
          is {\it continuous}. For now, we insist that the expression
          \b \Tr(U[k,x]\,\ss)\not=0\;, \e
          for all $(k,x)\in{\mathbb R}^2$; later, we briefly discuss a relaxation of
          this condition, but we will always require that
          $\Tr(\ss)\not=0$.  The operator $\ss$ will allow us to
          generalize the discussion of the previous section. As an
          advance notice we point out that if we make the special choice
          that
            \b \ss=|\s 0\>\s\<0| \;, \e
            then the general discussion that follows refers to the case of the Husimi --
            Glauber-Sudarshan dual pairs.

            We begin again with the symmetric expression for
            $\Tr(A^\dag\s B)$ given by (\ref{w15})
            which we modify so that
            \b && \hskip2.5cm\Tr(A^\dag\s B)\nonumber\\ &=& \int \frac{{\tilde A}[k,x]^*}
            {\Tr(U[k,x]\s\ss)} \,\{\Tr(U[k,x]\s\ss){\tilde
            B}[k,x]\}\,\frac{dk\s dx}{2\pi}\nonumber\\
&=&\int \{\frac{{\tilde A}[k,x]}{\Tr(U[k,x]^\dag\s\ss^\dag)}\s\}^*\, \{\s\Tr(U[k,x]\s\ss){\tilde
            B}[k,x]\s\}\,\frac{dk\s dx}{2\pi} \nonumber \\
 &&\hskip1.5cm\equiv \int {\tilde A}_{-\ss}[k,x]^*\,{\tilde B}_\ss[k,x]\,\frac{dk\s dx}{2\pi}\nonumber\\
&&\hskip1.5cm\equiv \int A_{-\ss}[p,q]^*\,B_\ss[p,q]\,\frac{dp\s
dq}{2\pi}\;. \e
    In the final line we have introduced the Fourier
    transform of the symbols in the line above. Our task is to find
    alternative expressions involving the symbols $A_{-\ss}[p,q]$
    and $B_\ss[p,q]$ directly in their own space of definition
    rather than implicitly through a Fourier transformation.

    We begin first with the symbol $B_\ss[p,q]$. In particular,  we
    note that
     \b &&B_\ss[p,q]=\int e^{i(kq-xp)}\,\Tr(U[k,x]\s\ss){\tilde
            B}[k,x]\,\frac{dk\s dx}{2\pi}\nonumber\\
&&= \int\Tr(U[p,q]^\dag\,U[k,x]\,U[p,q]\s\ss)
       \,\Tr(U[k,x]^\dag\s B)\,\frac{dk\s dx}{2\pi}\nonumber\\
&&= \int \Tr(U[k,x]\,U[p,q]\s\ss\,U[p,q]^\dag)
       \,\Tr(U[k,x]^\dag\s B)\,\frac{dk\s dx}{2\pi}\nonumber\\
       &&\hskip1.4cm=\Tr(U[p,q]\s\ss\,U[p,q]^\dag\s B)\;, \e
       where in the middle line we have used the Weyl form of the commutation
       relations Eq.~(\ref{u9}), and in the last line we have used Eq.~(\ref{eq:basic}), which
       has led us to the desired expression for $B_\ss[p,q]$. This
       expression is the sought for
       generalization of the Husimi representation; indeed, if $\ss=|\s 0\>\s\<0|$
       it follows immediately that
         \b B_{\ss}[p,q]&=&\Tr(U[p,q]\s|\s 0\>\s\<0|\s
         U[p,q]^\dag\s B)\nonumber\\ &=&\<p,q|\s B\s|p,q\>=B_H[p,q]\;.\e

     For general $\ss$, to find the expression for $A_{-\ss}[p,q]$ we appeal to the
     relation
        \b &&\Tr(A^\dag\s B)=\int A_{-\ss}[p,q]^*\,B_\ss[p,q]\frac{dp\s dq}{2\pi}\nonumber \\
&&=\int A_{-\ss}[p,q]^*\Tr(U[p,q]\s\ss\,U[p,q]^\dag\s B)\frac{dp\s
dq}{2\pi}\;, \e
        an equation which, thanks to its validity for all $B$ of the
        form $B=|\phi\>\s\<\psi|$ for arbitrary $|\phi\>$ and $|\psi\>$ in the Hilbert space,
         carries the important implication that
          \b A^\dag\equiv \int A_{-\ss}[p,q]^*\,U[p,q]\s\ss\,U[p,q]^\dag\,\frac{dp\s dq}{2\pi}\;, \e
        or if we take the Hermitian adjoint that
   \b A\equiv \int A_{-\ss}[p,q]\,U[p,q]\s\ss^\dag\,U[p,q]^\dag\,\frac{dp\s dq}{2\pi}\;.\label{u10} \e
        Observe that this equation implies a very general operator
        representation as a linear superposition of basic operators
        given by
            $U[p,q]\s\ss^\dag\,U[p,q]^\dag $,
            for a general choice of $\ss$ that satisfies the
            conditions given initially.

        Equation (\ref{u10}) for $A$ is the sought for generalization of the
        Glauber-Sudarshan representation; indeed, if $\ss=|\s
        0\>\s\<0|$,
        it follows immediately that
          \b A&=&\int A_{-\ss}[p,q]\,U[p,q]\s|\s 0\>\s\<0|\s U[p,q]^\dag\,\frac{dp\s dq}{2\pi}\nonumber\\
&=&\int A_{-\ss}[p,q]\;|p,q\>\s\<p,q|\,\frac{dp\s dq}{2\pi}\\
&=&\int A_{G-S}[p,q]\;|p,q\>\s\<p,q|\,\frac{dp\s dq}{2\pi}\;. \e

   Once again there is a direct connection between the
   generalization of the
   Husimi representation, $A_\ss[p,q]$, and the generalization of the
    Glauber-Sudarshan representation, $A_{-\ss}[p,q]$. In
    particular, it follows from (\ref{u10}) that
    \b A_\ss[r,s]&=&\no\\
    &&\hskip-1.6cm=\int A_{-\ss}[p,q]\,\Tr(U[r,s]\s\ss U[r,s]^\dag U[p,q]\s\ss^\dag
    U[p,q]^\dag)\,\frac{dp\s dq}{2\pi}\no\\
      &&\hskip-1.6cm=\int A_{-\ss}[p,q]\s\no\\
      &&\hskip-1cm\times[\Tr(U[r-p,q-s]\s\ss U[r-p,q-s]^\dag\ss^\dag)
      \,\frac{dp\s dq}{2\pi}\,. \label{q9}\e
      Again, this equation is a convolution, which just reflects the
      multiplicative connection between these two symbols in the
      Fourier space. Note that the convolution kernel in (\ref{q9}) is generally
      complex unless $\ss^\dag=\ss$; c.f., Eq.~(\ref{q8}).

\section{Examples of Generalized Phase Space Operator Representations}
In this section we offer a sample of the generalization offered by
our formalism. In particular, let us choose a thermal density matrix
  \b \ss=Z\,e^{-\beta N}\;, \e
  where $\beta>0$, $N=a^\dag a$ is the number operator, with
  spectrum $\{0,1,2,3,\dots\}$, and $Z=1-e^{-\beta}$ normalizes
  $\ss$ so that $\Tr(\ss)=1$. Observe that if we take a limit in
  which
  $\beta\ra\infty$, then $\ss\ra|\s 0\>\s\<0|$ appropriate to the
  standard case. For convenience, we label this example  simply by
  the parameter $\beta$.

  It follows first that
    \b \Tr(U[k,x]\s\ss)=Z\,\Tr(U[k,x]\s e^{-\bb
    N})=e^{-\t(k^2+x^2)/4}, \e
    where
      \b \t\equiv\frac{1+e^{-\bb}}{1-e^{-\bb}}\;. \e
      Clearly, $\t>1$, and $\t\ra1$ as $\bb\ra\infty$.
      Furthermore,
    \b &&B_\bb[p,q]=Z\,\Tr(U[p,q]\s e^{-\bb N}\s U[p,q]^\dag\s B)\nonumber\\
&&=(1-e^{-\bb})\sum_{n=0}^\infty\,e^{-\bb n}\<\s n|U[p,q]^\dag\s B(P,Q)\s U[p,q]\s|n\>\nonumber \\
&&=(1-e^{-\bb})\sum_{n=0}^\infty\,e^{-\bb n}\<\s n|\s B(P+p,Q+q)\s|n\> \;.\e

      This expression admits alternative forms as well. Based on
      the relation
        \b {\tilde B}_\bb[k,x]=e^{-\t(k^2+x^2)/4}\,{\tilde B}[k,x]\;,
        \e
        it follows that
        \b B_\bb[p,q]=\int
        e^{-(1/\t)[(p-p')^2+(q- q')^2]}\frac{B[p',q']}{\pi\s\t}\frac{dp' dq'}{2\pi}\;.
        \e
        Still another connection is given by
         \b &&B_\bb[p,q]=\int e^{i(kq-xp)}\,e^{-\t(k^2+x^2)/4}\,{\tilde
         B}[k,x]\,\frac{dk\s dx}{2\pi}\nonumber \\
&&=e^{(\t/4)(\d^2/\d p^2+\d^2/\d q^2)}\int e^{i(kq-xp)}\,{\tilde
         B}[k,x]\,\frac{dk\s dx}{2\pi}\nonumber \\
&&=e^{(\t/4)(\d^2/\d p^2+\d^2/\d q^2)}\,B[p,q]\;. \e

 Let us turn our attention to the dual representation
 associated with $A_{-\bb}[p,q]$. In this case, we focus on the
 operator representation given by
   \b &&A = (1-e^{-\bb})\,\int A_{-\bb}[p,q]\,U[p,q]\s e^{-\bb N}\s
   U[p,q]^\dag \,\frac{dp\s dq}{2\pi}\nonumber\\
&&\hskip.38cm= \,(1-e^{-\bb})\sum_{n=0}^\infty e^{-\bb n}\s
   \int A_{-\bb}[p,q]\no\\
   &&\hskip2.5cm\times\,|p,q;n\>\<p,q;n|\frac{dp\s dq}{2\pi}\;,  \e
where we have introduced the notation \cite{nstates}
  \b |p,q;n\>\equiv U[p,q]\s|n\> \e
  for those coherent states for which the fiducial vector is $|\s
  n\>$. There is an alternative representation for the symbol under
  discussion which is given by
     \b A_{-\bb}[p,q]=e^{-(\t/4)(\d^2/\d p^2+\d^2/\d
         q^2)}\,A[p,q]\;,  \e
         which is particularly useful if the Weyl symbol is
         a polynomial.

  We now turn our attention to a different but closely related
  example, which we distinguish by a prime ($'$). In particular, we
  choose
  \b \ss'\equiv Z'\,(-1)^N\,e^{-\bb N}=(1+e^{-\bb})\,e^{-(\bb+i\pi)\s
  N}\;. \e
  This example also corresponds to a Hermitian choice for $\ss$, but
  it is not positive definite as was our previous choice. The
  results of interest are easily calculated simply by an analytic
  extension of the parameter $\bb$ to $\bb'\equiv\bb+i\pi$. In particular,
  it follows that
     \b \Tr(U[k,x]\,\ss')=e^{-\t'(k^2+x^2)/4}\;, \e
     where
      \b
      \t'=\frac{1+e^{-\bb'}}{1-e^{-\bb'}}=\frac{1-e^{-\bb}}{1+e^{-\bb}}\;.
      \e
      Thus, in the present case, we have a similar rescaling factor, but
      this time, the parameter $\t'<1$ , with again the limit that $\t'\ra1$ as
      $\bb\ra\infty$.

      Turning attention to the relevant symbols, we first see that
      \b &&B_{\bb'}[p,q]=Z'\,\Tr(U[p,q]\s e^{-\bb' N}\s U[p,q]^\dag\s B)\nonumber\\
&&\hskip-.3cm=(1+e^{-\bb})\sum_{n=0}^\infty\,(-1)^n\,e^{-\bb n}\<\s n|U[p,q]^\dag\s B(P,Q)\s U[p,q]\s|n\>\nonumber\\
&&\hskip-.3cm=(1+e^{-\bb})\sum_{n=0}^\infty\,(-1)^n\,e^{-\bb n}\<\s n| B(P+p,Q+q)|n\> \;.\e
     This expression also admits alternative forms as well. Based on
      the relation
        \b {\tilde B}_{\bb'}[k,x]=e^{-\t'(k^2+x^2)/4}\,{\tilde B}[k,x]\;,
        \e
        it follows that
        \b B_{\bb'}[p,q]=\int
        e^{-(1/\t')[(p-p')^2+(q- q')^2]}\frac{B[p',q']}{\pi\s\t'}\frac{dp'dq'}{2\pi}\;.\nonumber\\
        \e
        Still another connection is given by
         \b B_{\bb'}[p,q]=e^{(\t'/4)(\d^2/\d p^2+\d^2/\d
         q^2)}\,B[p,q]\;. \e

         Let us again turn our attention to the dual representation
 associated with $A_{-\bb'}[p,q]$. In this case, we focus on the
 operator representation given by
   \b &&A = (1-e^{-\bb'})\,\int A_{-\bb'}[p,q]\,U[p,q]\s e^{-\bb' N}\s
   U[p,q]^\dag \frac{dp\s dq}{2\pi}\nonumber\\
&&\hskip.38cm=(1+e^{-\bb})\sum_{n=0}^\infty\,(-1)^ne^{-\bb n}\,
   \int A_{-\bb'}[p,q]\no\\
   &&\hskip2.7cm\times\,|p,q;n\>\s\<p,q;n|\frac{dp\s dq}{2\pi}\;. \nonumber\\  \e
    There is another representation for the symbol under
  discussion given by
     \b A_{-\bb'}[p,q]=e^{-(\t'/4)(\d^2/\d p^2+\d^2/\d
         q^2)}\,A[p,q]\;,  \e
         which is particularly useful, once again, if the Weyl symbol is
         a polynomial.

         It is of interest to observe, for the thermal choice of $\ss$ (or
         one with a complex temperature such as $\ss'$) discussed in this section, that
         every bounded operator $A$ may be
         represented for any choice of $\bb$ by weights $A_{-\bb}[p,q]$
         that are elements of ${\cal Z}'({\mathbb R}^2)$, just as was
         the case for the Glauber-Sudarshan representation. This
         holds because if we multiply a distribution in ${\cal
         D}'({\mathbb R}^2)$ by an expression of the form
         $\exp[\t(k^2+x^2)/4]$, it remains a distribution in ${\cal
         D}'({\mathbb R}^2)$ because multiplication with such a factor
         leaves the test function space ${\cal D}({\mathbb R}^2)$ invariant.

         The example discussed above has some similarity with one
         treated by Cahill and Glauber \cite{cg}. In their work, they
         interpret the additional factor as arising from a
         reordering of the basic Weyl operator, i.e., a smooth transition from
         normal ordered to anti-normal ordered. On the other
         hand, we interpret a similar factor as coming from an
         alternative definition of the associated symbol.

\section{Further Examples and Generalizations}
\subsection{Additional examples}
For a fixed, nonzero point $(r,s)$ in phase space, consider the
non-Hermitian example for which
  \b \ss=(1-e^{-\bb})\,U[r,s]^\dag\,e^{-\bb\s N}\;.  \e
  In this case
    \b &&\Tr(\s
    U[k,x]\s\ss)=(1-e^{-\bb})\,\Tr(\s U[k,x]\,U[r,s]^\dag\,e^{-\bb\s
    N}\s)\nonumber\\
&&=(1-e^{-\bb})\,e^{-i(ks-xr)/2}\s\Tr(\s U[k-r,x-s]\,e^{-\bb\s N}\s)\nonumber\\
&&=e^{-i(ks-xr)/2}\s e^{-\t[(k-r)^2+(x-s)^2]/4\s}\;,  \e
where
     \b \t \equiv\frac{1+e^{-\bb}}{1-e^{-\bb}}  \e
   as before. With this example, we find that
   \b &&\hskip2.00cm {\tilde B}_\ss[k,x] = \nonumber\\&&e^{-i(ks-xr)/2}\s e^{-\t[(k-r)^2+(x-s)^2]/4\s}\,\Tr(\s
   U[k,x]^\dag\s B)\;,~~~~~~~~ \e
   which has the feature that phase space points at and near $(r,s)$
   are ``emphasized'' in comparison with the rest of phase space. For
   applications in the time-frequency domain, for example, such an emphasis may
   of value.

   Another class of examples is offered by the following observation.
   First, let us make a preliminary choice of $\ss=|n\>\s\<n|$,
   where $|n\>$ denotes the
   $n$ particle state introduced earlier. For such a choice we learn
   that (see, e.g., \cite{cg})
    \b &&\Tr(\s U[k,x]\s|n\>\s\<n|)= \<n|\s U[k,x]\s|n\>~~~~~~~\nonumber\\ &&={\rm
    L}_n(\half(k^2+x^2))\,e^{-(k^2+x^2)/4}\;,  \e
    where ${\rm L}_n(y)$ denotes a Laguerre polynomial defined, as
    usual, by
      \b {\rm L}_n(y)\equiv\frac{e^y}{n!}\s\frac{d^n}{dy^n}\s( y^n\s
      e^{-y})\;.  \e
      Each Laguerre polynomial ${\rm L}_n(y)$  is real and has $n$
      distinct, real zeros. Thus the choice $\ss=|n\>\s\<n|$ violates the basic
      postulate that $\Tr(U[k,x]\s\ss)\not=0$ for all $(k,x)\in
      {\mathbb R}^2$ whenever $n>0$. However, we can overcome this problem by
      choosing instead a non-Hermitian example, such as
        \b \ss=R\,|n\>\s\<n|+i\s T\,|m\>\s\<m| \;, \e
        where $m\not=n$, and $R$ and $T$ are two real, nonzero factors
        that reflect the relative weight of
    the two contributions.
        As a consequence, we are led to
         \b && \Tr(\s U[k,x]\s\ss) =[\s R\,{\rm L}_n(\half(k^2+x^2))\nonumber\\
         &&\hskip1cm+~i\s T\,{\rm L}_m(\half(k^2+x^2))\s]\,e^{-(k^2+x^2)/4}\;, \e
         which is never zero for any choice of $(k,x)$.
Further extensions along these lines are evident.

Associated with each choice of $\ss$ above is the corresponding dual
representation given, as usual, by
  \b A\equiv \int A_{-\ss}[p,q]\,U[p,q]\s\ss^\dag\,U[p,q]^\dag\s\frac{dp
        dq}{2\pi}\;, \e
which we do not need to spell out in detail again.

\subsection{Relaxation of nonvanishing criterion}
 Returning to a general choice of $\ss$, we observe
that so long
   as $\Tr(U[k,x]\s\ss)$ never vanishes, we can assert that if
   $B_\ss[p,q]=0$ for all $(p,q)\in{\mathbb R}^2$, then
   the operator $B=0$. This property follows from the
   fact that when ${\tilde B}[k,x]=0$, it follows that
    $\Tr(B^\dag\s B)=\tint |{\tilde B}[k,x]|^2\,dk\s
   dx=0$, which can only happen if $B=0$. However, if
   \b \Tr(U[k_o,x_o]\s\ss)=0  \e
for some phase space point $(k_o,x_o)$, for example, then the very
operator $B=U[k_o,x_o]$ would lead to the fact that ${\tilde
B}[k,x]=0$, for all $(k,x)$ in phase space, with the consequence
that $B_\ss[p,q]\equiv0$ despite the fact that $B\not=0$.

   Turning to the dual symbols, we observe that so long
   as $\Tr(U[k,x]\s\ss)$ never vanishes, we can formally represent every
   operator in the form
   \b  A=\int A_{-\ss}[p,q]\,U[p,q]\s\ss^\dag\s
   U[p,q]^\dag\s\frac{dp\s
        dq}{2\pi}, \e
   where the symbol
   \b   A_{-\ss}[p,q]=\int
   e^{i(px-qk)}\,\frac{{\tilde
   A}[k,x]}{\Tr(U[k,x]^\dag\s\ss^\dag)}\,\frac{dk\s
   dx}{2\pi} \e
is a distribution appropriate to
   the situation under consideration. More precisely, we can represent all
   bounded operators in such a fashion, and every operator can be obtained as a
   suitable limit from the set of bounded operators.
 On the other hand, if we relax the condition that $\Tr(U[k,x]\s\ss)$
   never vanishes, we loose this generality. Suppose again that
     $\Tr(U[k_o,x_o]\s\ss)=0 $
     for some particular point in phase space $(k_o,x_o)$. In that
     case, it is strictly speaking not possible to generate the operator
       $ U[k_o,x_o]\s\ss^\dag\s U[k_o,x_o]^\dag\;  $.

   On the other hand, if one is only interested in operators that
   are {\it polynomial} in $P$ and $Q$, then the support of the Weyl
   symbol ${\tilde A}[k,x]$ or ${\tilde B}[k,x]$ is strictly at the
   origin in phase space, e.g., for ${\tilde C}$ equal either
   ${\tilde A}$ or ${\tilde B}$,

     \b {\tilde C}[k,x] =\sum_{i,j=0}^{I,J}
     c_{i,j}\frac{d^i}{dk^i}\,\frac{d^j}{dx^j}\,\delta(k)\,\delta(x)\;,
     \e
     where $I<\infty$ and $J<\infty$. In that case, the fact that
     $\Tr(U[k,x]\s\ss)$ may vanish away from the origin has no
     influence on the matter, and thus so long as $\Tr(\ss)\not=0$,
     and thus by continuity $\Tr(U[k,x]\s\ss)$ is nonzero in an open neighborhood of the
     origin,
     {\it all} polynomials in $P$ and $Q$ can be represented in the form
       \b A=\int A_{-\ss}[p,q]\,U[p,q]\s\ss^\dag\s U[p,q]^\dag\frac{dp\s dq}{2\pi}\;, \e
       and, additionally, the symbol
       \b B_\ss[p,q]=\Tr(U[p,q]\s\ss\s U[p,q]^\dag\s B)\e
       uniquely determines the operator $B$ provided that it is a
       polynomial.
   This result generalizes a result
   established some time ago \cite{klalong}.

   \subsection{Arbitrary weight factors}
   It is of course possible to reweight the Weyl
   representation by quite arbitrary factors in the manner
     \b &&\Tr(A^\dag\s B)=\int {\tilde A}[k,x]^*\,{\tilde
     B}[k,x]\frac{dk\s dx}{2\pi}\nonumber \\
&&=\int\{\s (F[k,x]^*)^{-1}\,{\tilde A}[k,x]\}^*\,\{\s
     F[k,x]\,{\tilde B}[k,x]\}\frac{dk\s dx}{2\pi}\nonumber \\
&&\equiv \int {\tilde A}_{-F}[k,x]^*\,{\tilde
      B}_F[k,x]\frac{dk\s dx}{2\pi} \nonumber\\
&&\equiv \int A_{-F}[p,q]\,B_F[p,q]\frac{dp\s dq}{2\pi}\;,
      \e
      where $F[k,x]$ is an arbitrary nonvanishing factor,
      and, as usual, we have introduced the Fourier transform
      elements in the last line. Apart from additional conditions
       -- specifically, $F(0,x)=1=F(k,0)$ -- this is the procedure used by
      Cohen \cite{cohen}.
      However, a general factor $F[k,x]$ cannot normally be
      represented as $\Tr(\s U[k,x]\s\ss)$ for some $\ss$.
      For example, the function $\Tr(\s U[k,x]\s\ss)$ is necessarily
      continuous and bounded by $|\s\Tr(\s U[k,x]\s\ss)\s|\le
      \Tr(\sqrt{\ss^\dag\s\ss}\s)$, the trace class norm of $\ss$.
      However, more to the point, the introduction of a general
      expression such as $F[k,x]$ would normally introduce elements
      outside the Hilbert space and its operators that we have so
      far consistently stayed within. Such an extension may of
      course be considered, but the non-operator nature in the extension being made
      should be appreciated and accepted.

      Of course, the investigation of quasi-distributions continues
      unabated. As one comparatively recent example of such investigations,
      we cite the work of \cite{sudar}.\vskip0cm

      \subsection{Non-canonical generalizations}
      Representations of Hilbert space operators in the manner of
      the Weyl representation may be carried
      out for a great variety of groups. In addition, since coherent states may be defined
      for other groups, analogs of the Husimi
      and dual Glauber-Sudarshan representations exist in such cases as well, and these have
      often been discussed in the literature.
      Consequently, it follows that asymmetric representations of
      various forms, analogous to those presented in this paper for the Weyl group, can
      be introduced for other groups, e.g., the groups SU($2)$ and SU($1,1$), as well.

      \section{CONCLUSIONS}
      Quite naturally, an increase in the family of representations
      of various systems offers new ways to study such systems. A
      general formulation of states and observables must lead to a
      symmetric, abstract description; but for {\it specific}
      systems, where the states of interest may be ``better
      behaved''
      than the observables, it may be useful to take advantage of that
      distinction with an asymmetric representation pair. Likewise,
      within the realm of time-frequency analysis, signals and their
      analyzers may have very different characteristics that also
      suggest the utility of an asymmetric pair. Moreover, to
      minimize any extraneous elements in such asymmetric
      representations, it is appropriate that the representations in
      question be formulated within the initial abstract
      formalism, a point of view developed in the present paper.

      Although it is difficult to predict in advance just which
      asymmetric representation pairs may prove useful, there may
      well arise special situations for which certain asymmetric
      representations prove useful -- just as was the case in quantum optics for
      the Husimi -- Glauber-Sudarshan representation pair.

\section*{ACKNOWLEDGEMENTS}
JRK is pleased to acknowledge the hospitality of the Complex Systems
and Soft Materials Research Group of the Department of Physics, The
Norwegian University of Science and Technology, in Trondheim, as
well as the support provided through the 2006 Lars Onsager
Professorship for an extended visit during which time this paper was
prepared.

\end{document}